\documentclass[%
 reprint,
superscriptaddress,prl,
 amsmath,amssymb,
 aps,
]{revtex4-2}

\usepackage{graphicx}
\usepackage{dcolumn}
\usepackage{bm}

\newcommand{\BSMFO}{Ba$_{0.4}$Sr$_{1.6}$Mg$_2$Fe$_{12}$O$_{22}$}
\newcommand{\kZ}{k_{0}}
\newcommand{\kT}{k_{2}}
\newcommand{\kTh}{k_{3}}
\newcommand{\rlu}{\text{r.l.u.}}
\newcommand{\Phat}{\hat{\bm{P}}}
\newcommand{\Qhat}{\hat{\bm{Q}}}
\newcommand{\Mperp}{\bm{M}_{\perp}}

\begin{document}


\title{Phase-Coherent Spin and Lattice Modulations in a Y-type Hexaferrite}

\author{O. Stepančić}
\affiliation{Division of Synchrotron Radiation Research, Department of Physics,
Lund University, SE-22100 Lund, Sweden}

\author{A. Stellhorn}
\affiliation{European Spallation Source ERIC, Partikelgatan 2, 224 84 Lund, Sweden}

\author{E. Campillo}
\affiliation{Division of Synchrotron Radiation Research, Department of Physics,
Lund University, SE-22100 Lund, Sweden}

\author{K. Zhai}
\affiliation{Center for High Pressure Science, State Key Laboratory of Metastable Materials
Science and Technology, Yanshan University, Qinhuangdao, China}

\author{N. Ekström}
\affiliation{European Spallation Source ERIC, Partikelgatan 2, 224 84 Lund, Sweden}

\author{A. Corani}
\affiliation{European Spallation Source ERIC, Partikelgatan 2, 224 84 Lund, Sweden}

\author{D. Alba Venero}
\affiliation{ISIS neutron and muon source, STFC, Rutherford Appleton Laboratory, Didcot OX11 0QX, United Kingdom}

\author{R. Cubitt}
\affiliation{Institut Laue--Langevin, CS 20156, 38042 Grenoble Cedex 9, France}

\author{E. Blackburn}
\email{elizabeth.blackburn@fysik.lu.se}
\affiliation{Division of Synchrotron Radiation Research, Department of Physics,
Lund University, SE-22100 Lund, Sweden}

\author{L. Shen}
\email{lingjia.shen@fysik.lu.se}
\affiliation{Division of Synchrotron Radiation Research, Department of Physics,
Lund University, SE-22100 Lund, Sweden}


\begin{abstract} 
Spin-driven multiferroics owe their electric polarization to lattice distortions that follow the magnetic order, yet in incommensurate magnets symmetry normally forbids such a distortion at the fundamental magnetic wavevector, and the distortions themselves are rarely observed. Here we use half-polarized neutron scattering to show that the Y-type hexaferrite \BSMFO\ hosts, in its saturated high-field state, a magnetic modulation and a longitudinal lattice modulation at one and the same wavevector, locked in phase. The signature is a large nuclear-magnetic interference asymmetry, $r\simeq0.8$, of the same sign at the two conjugate satellites and persisting to $275$~K. The zone contrast and the field dependence of the satellite indicate an internal, field-induced longitudinal distortion. The modulated state coexists with the transverse-conical magnetoelectric phase and outlives it, revealing a fundamental-wavevector spin-lattice coupling in a magnetic field regime previously regarded as a simple ferrimagnet.
\end{abstract}

\maketitle

\maketitle

In spin-driven multiferroics, improper ferroelectricity is encoded in the lattice~\cite{Kimura2003,Kenzelmann2005,Cheong2007,kimura-arcmp-2012,Fiebig2016,Spaldin2019}: magnetic order lowers the crystal symmetry and can induce an atomic displacement pattern that carries ferroelectric polarization. Although magnetic structures can often be determined with high precision, the accompanying displacements -- typically only femtometers in amplitude~\cite{Walker2011,BrownChatterji2013} -- are rarely observed directly and are instead inferred from the polarization they are expected to generate.

Symmetry places a further constraint on these displacements. In a single-$k$ incommensurate magnetic state without a uniform magnetic component, the structural response at the fundamental magnetic wavevector is forbidden and the leading lattice modulation occurs at $2k$. Formally, the superspace operation $(1'|000,\tfrac{1}{2})$ restricts the magnetic modulation to odd and the structural modulation to even Fourier harmonics~\cite{PerezMato2012,Jia2007}. A uniform moment $M_0$ lifts this restriction through cross terms of the form $M_0M_{\bm q}u_{\bm q}$~\cite{Aliouane2006,Jia2007}, but a lattice modulation at the fundamental wavevector has been resolved in only a few cases~\cite{Walker2011,Aliouane2006,Strempfer2008,Terada2008,Arima2007}. In multiferroic hexaferrites, some diffraction studies have reported a lattice modulation at the fundamental magnetic wavevector~\cite{Asaka2011,Ueda2016}, while a resonant X-ray study of a related composition reached the opposite conclusion~\cite{Chmiel2019}; the relative phase between the nuclear and magnetic modulations has not previously been measured.

Polarized neutrons provide a direct, phase-sensitive test because a satellite carrying both nuclear and magnetic structure factors produces a polarization-dependent interference term. Using this technique we report a high-field state in \BSMFO\ in which a field-parallel magnetic modulation interferes coherently with a longitudinal lattice modulation at the same fundamental wavevector, beyond the field range of the well established spin-current-driven magnetoelectric phase~\cite{kimura-arcmp-2012}.

\begin{figure*}[t] 
\includegraphics[width=\textwidth]{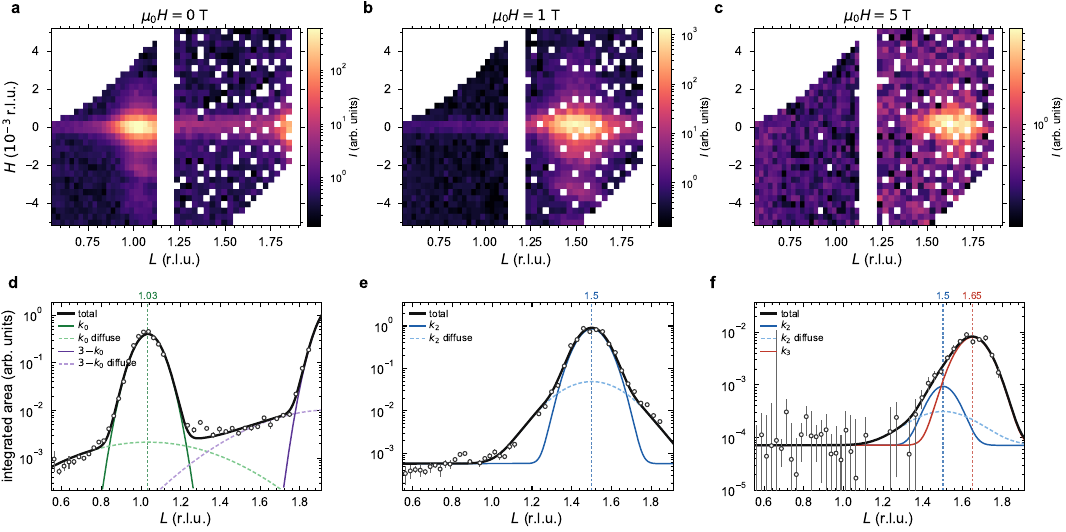} 
\caption{Field evolution of the $(0,\,0,\,L)$ scattering at $5$~K in the $I^{+}$ channel. (a)--(c) Reciprocal-space maps measured at $0$, $1$, and $5$~T. (d)--(f) Corresponding longitudinal profiles and fits. Vertical guides indicate $L=1.5$, $L=1.65$, and the fitted center of the $\kZ$ satellite. With increasing field, the zero-field incommensurate $\kZ$ satellite gives way to the commensurate $\kT$ component, while a third satellite, $\kTh$, becomes clearly resolved near $L=1.65$ above approximately $2$~T.} 
\label{fig:1} 
\end{figure*}

For half-polarized neutrons, the cross section contains two polarization-dependent terms~\cite{Blume1963,Maleev2002,Ressouche2014},
\begin{equation}
I^{\pm} = |N|^{2}+|\Mperp|^{2} \pm 2\Phat\!\cdot\!\mathrm{Re}(N^{*}\Mperp) \pm i\,\Phat\!\cdot\!(\Mperp^{*}\!\times\!\Mperp),
\label{eq:bm}
\end{equation}
where $N$ and $\Mperp$ are the nuclear and transverse magnetic structure factors; the two terms represent nuclear-magnetic interference and magnetic chirality. We characterize a satellite by the normalized asymmetry
\begin{equation}
r \equiv \frac{I^{+}-I^{-}}{I^{+}+I^{-}} = \frac{ 2\Phat\!\cdot\!\mathrm{Re}(N^{*}\Mperp) +i\Phat\!\cdot\!(\Mperp^{*}\!\times\!\Mperp) }{ |N|^{2}+|\Mperp|^{2} },
\label{eq:asymmetry}
\end{equation}
where $I^{\pm}$ are the integrated satellite intensities for the two incident polarization states and the second equality holds for ideal polarization. Finite polarization and flipper efficiency reduce the measured asymmetry by a factor $\mathcal{P}_{\mathrm{eff}}\leq1$; neglecting this correction makes the phase bound derived below conservative~\cite{SM}.

The two polarization-dependent terms have opposite parity under $\bm q\rightarrow-\bm q$. For a real-space structure, $N(-\bm q)=N^{*}(\bm q)$ and $\Mperp(-\bm q)=\Mperp^{*}(\bm q)$, so the interference contribution to $r$ is even under interchange of the two conjugate satellites whereas the chiral contribution is odd. In the present geometry the chiral term is, in addition, geometrically suppressed: for a satellite on the $(0,\,0,\,L)$ rod the vector $\Mperp^{*}\!\times\!\Mperp$ is parallel to $\Qhat\parallel\bm c^{*}$, and with $\Phat$ in the basal plane its projection onto $\Phat$ is at most $\sin\theta\simeq0.1$ of its magnitude at the $\it{Bragg}$ angle $\theta$ of the satellite. Any sizeable asymmetry on this rod therefore originates in nuclear-magnetic interference; the conjugate-satellite comparison checks this assignment and the reproducibility of the measurement.

We studied single-crystalline \BSMFO, a Y-type hexaferrite with space group $R\bar{3}m$, composed of alternating large- and small-moment ($L$ and $S$) magnetic blocks. This compound hosts a magnetoelectric transverse-conical phase under a small in-plane magnetic field~\cite{Kimura2005,Ishiwata2008,Zhai2017,Kitagawa2010}. Half-polarized small-angle neutron scattering was performed on the D33 instrument at the Institut Laue-Langevin using an incident wavelength of $\lambda=5$~\AA~and $\mathcal{P}_{\mathrm{eff}}$~$\approx$~0.94 (polarizer efficiency) \cite{ILLDATA,Dewhurst2016}. Complementary time-of-flight measurements were performed on the ZOOM instrument at the ISIS Neutron and Muon Source~\cite{ISISDATA,SM}. In both experiments, the magnetic field was applied in the basal plane along $[100]$, with the incident neutron polarization $\bm P\parallel\bm H$ and therefore $\hat{\bm P}\perp\bm c^{*}$. Reciprocal-space maps of the $(0,\,0,\,L)$ rod were reconstructed from sample rocking scans. Magnetic neutron scattering probes only the component $\Mperp=\bm M-(\bm M\!\cdot\!\Qhat)\Qhat$ perpendicular to the scattering vector. Along the $(0,\,0,\,L)$ direction, $\Qhat\parallel\bm c^{*}$, while the nuclear-magnetic interference term is proportional to $\hat{\bm P}\!\cdot\!\mathrm{Re}(N^{*}\Mperp)$. Because $\hat{\bm P}\parallel\bm H$, the interference selectively probes the field-parallel in-plane component of the magnetic modulation, $\hat{\bm P}\!\cdot\!\Mperp=M_{\parallel H}$.

Figure~\ref{fig:1} summarizes the field evolution of the $(0,\,0,\,L)$ scattering in the $I^{+}$ channel. At zero field, the scattering is dominated by the incommensurate $\kZ\approx(0,\,0,\,1.03)$ component. With increasing field, the commensurate $\kT=(0,\,0,\,1.5)$ component develops, followed at higher fields by an additional, much weaker satellite, denoted $\kTh$, at $L=1.646(6)$. The $\kZ$ and $\kT$ components have previously been identified as longitudinal-conical and twofold transverse-conical states~\cite{Zhai2017}; the present $\kZ$ position differs slightly from the reported $(0,\,0,\,0.96)$, which may reflect sample composition. The $\kTh$ satellite is the focus of this work.

The following numerical model on the complete field series is used to separate the overlapping contributions. The $\kZ$ and $\kT$ signals each require narrow and diffuse Gaussian components. The diffuse scattering may reflect short-range spin correlations induced by stacking disorder along the $c$ axis, which is well known in the block-stacked hexaferrite family~\cite{pullar-pms-2012}. The much weaker $\kTh$ satellite is adequately described by a single Gaussian within the available resolution and counting statistics. An $L$-independent term accounts for the residual rod scattering along $H=0$, likely associated with local defects. The total fits and their individual components are shown for three representative fields in Figs.~\ref{fig:1}(d)--\ref{fig:1}(f); further details of the fitting procedure and the corresponding $I^{-}$ data are provided in the Supplemental Material~\cite{SM}.

\begin{figure}[t]
\includegraphics[width=\columnwidth]{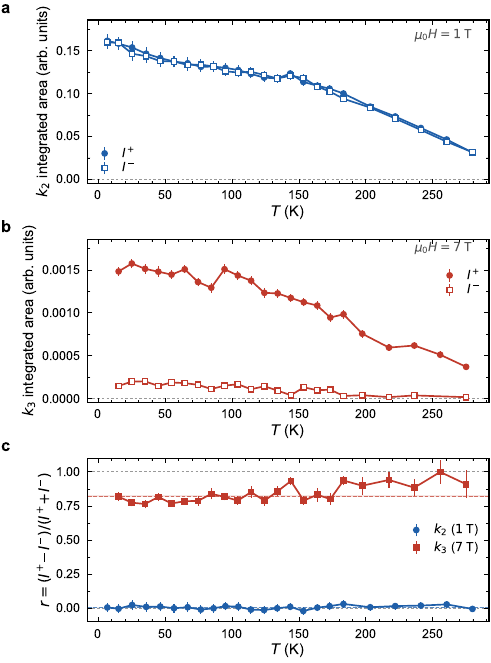}
\caption{Temperature dependence of the two satellites. (a) $\kT$ integrated area at $1$~T and (b) $\kTh$ at $7$~T, in both polarization channels. (c) Asymmetry $r$ of each. The commensurate $\kT$ satellite is polarization-independent, $r\simeq0$, whereas $\kTh$ maintains $r\simeq0.8$ at all temperatures probed. }
\label{fig:2}
\end{figure}

Figure~\ref{fig:2} compares the temperature dependence of $\kT$ at $1$~T with that of $\kTh$ at $7$~T. The integrated intensities of both satellites decrease monotonically upon warming, with the polarization-summed $\kTh$ intensity falling by a factor of $4.2$ between $15$ and $275$~K. Their polarization responses, however, are qualitatively different. The commensurate satellite remains polarization-independent, $r(\kT)=0.007(6)$ averaged over the measured temperature range with individual values within $\pm0.03$. Since the chiral term is geometrically suppressed on this rod, this null means $\mathrm{Re}[N^{*}(\kT)M_{\parallel H}(\kT)]\simeq0$ to within about $1~\%$ of the $\kT$ intensity: either the transverse-conical satellite carries no nuclear amplitude, or its modulated moment has no field-parallel component, as expected for a cone about $\bm H$~\cite{Zhai2017}; it does not constrain the chirality of $\kT$~\cite{SM}. Measured in the same scans with the same procedure, $\kT$ also provides an internal null for the polarization setup, normalization and fitting. By contrast, $r(\kTh)$ remains large, with a weak upward trend from $0.82(3)$ to $0.91(10)$, throughout the measured temperature range, even as the absolute satellite intensity is largely suppressed.

\begin{figure}[t]
\includegraphics[width=\columnwidth]{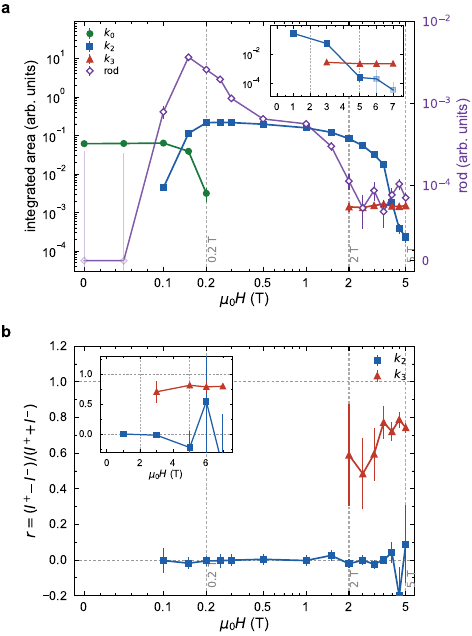}
\caption{Field dependence at $5$~K, $I^{+}$ channel. (a) Integrated areas of $\kZ$, $\kT$ and $\kTh$, with the flipper-independent rod intensity on the right axis; the field axis is nonuniform and reference lines mark $0.2$, $2$ and $5$~T. (b) Asymmetry parameters [Eq.~\eqref{eq:asymmetry}]. Insets: high-field data from a second set of measurements. Error bars exceeding the plotted value are shown faded.}
\label{fig:3}
\end{figure}

To place this polarization response in context, we next examine the field evolution of $\kZ$, $\kT$, $\kTh$, and magnetic defects [Fig.~\ref{fig:3}(a)]. The $\kZ\rightarrow\kT$ transition occurs near $0.2$~T. A comparatively narrow $\kTh$ satellite becomes clearly identifiable above approximately $2$~T and initially coexists with $\kT$. At lower fields, the $\kTh$ position overlaps the diffuse tail of the much stronger $\kT$ component; thus, $2$~T marks the lowest field at which a distinct $\kTh$ satellite is resolved within the sensitivity of the present measurements. By this field, the uniform magnetization has reached approximately $87\%$ of the extrapolated ferrimagnetic moment, $M_{\mathrm{sat}}=9.15(1)\,\mu_{\mathrm B}$/f.u., obtained from the zero-field intercept of a linear fit to the high-field magnetization (see Supplemental Material~\cite{SM}). This separation of field scales is noteworthy because a field-enabled magnetoelastic coupling at the fundamental wavevector can contain a cross term, $M_{0}M_{q}$, between the uniform and modulated magnetic components \cite{Aliouane2006,Jia2007,Walker2011}. The $L$-independent rod intensity (defects) also shows changes near the development of $\kT$ at $0.2$~T and the first resolved $\kTh$ signal at approximately $2$~T.

At higher fields, the $\kT$ intensity decreases by several orders of magnitude and becomes indistinguishable from zero within uncertainty by $6$~T [inset of Fig.~\ref{fig:3}(a)]. The $\kTh$ satellite, by contrast, remains well defined to at least $7$~T, extending beyond the field range in which the transverse-conical $\kT$ order is detectable. At this field, the dominant uniform ferrimagnetic component is saturated~\cite{SM}. The evolution from coexistence with $\kT$ to a high-field regime containing only the incommensurate $\kTh$ satellite identifies $\kTh$ as a distinct modulated state, rather than a subsidiary feature of the transverse-conical phase. Its survival in the ferrimagnetic saturation regime further demonstrates that the high-field state is not a simple uniform ferrimagnet, as has been proposed for materials of this class.

\begin{figure}[t]
\includegraphics[width=\columnwidth]{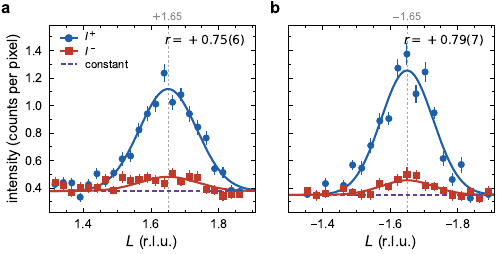}
\caption{The $\pm\kTh$ pair at $7$~T and $5$~K, reconstructed from separate detector panels. (a) $+\kTh$ and (b) $-\kTh$; note the mirrored abscissa. Curves are a joint fit of a Gaussian plus a polarization-independent constant. The asymmetry has the \emph{same} sign at both satellites, $r=+0.75(6)$ and $+0.79(7)$: the polarization dependence is dominated by the even nuclear-magnetic interference term, with no odd chiral contribution resolved [Eq.~\eqref{eq:parity}].}
\label{fig:4}
\end{figure}

The interference origin of the $\kTh$ asymmetry is confirmed by measuring the two conjugate satellites on separate detector panels [Fig.~\ref{fig:4}]. At $7$~T,
\begin{equation}
r(+\kTh)=+0.75(6),\qquad
r(-\kTh)=+0.79(7).
\end{equation}
Decomposing into components of definite parity gives
\begin{align}
r_{\mathrm{even}} &= \frac{1}{2} \left[ r(+\kTh)+r(-\kTh) \right] =+0.77(5), \notag\\ r_{\mathrm{odd}} &= \frac{1}{2} \left[ r(+\kTh)-r(-\kTh) \right] =-0.02(5).
\label{eq:parity}
\end{align}

The asymmetry is even under interchange of the conjugate satellites, as required for nuclear-magnetic interference, and its odd component is consistent with zero. Because the chiral channel is geometrically closed on this rod, the parity test checks the interference assignment and the reproducibility across panels rather than bounding the intrinsic chirality, which it does not constrain in this geometry. It follows that
\begin{equation}
\begin{gathered}
N(\kTh)\neq0,\qquad M_{\parallel H}(\kTh)\neq0,\\ \mathrm{Re}\!\left[ N^{*}(\kTh)M_{\parallel H}(\kTh) \right]\neq0.
\end{gathered}
\label{eq:chain}
\end{equation}
Thus the $\kTh$ satellite carries a nuclear structure factor that interferes coherently with the field-parallel magnetic component, at every field above $2$~T and every temperature [Figs.~\ref{fig:2}(c), \ref{fig:3}(b)]. A nonzero $M_{\parallel H}(\kTh)$ implies that the field-parallel moment itself is modulated at $\kTh$, a longitudinal, fan-like component that distinguishes the $\kTh$ state from a cone about $\bm H$; its $c$-axis component is not probed on this rod. Comparable nuclear-magnetic interference at the magnetic satellites of the commensurate phase of YMn$_2$O$_5$ has been used to infer displacements of order $0.05$~\AA~\cite{BrownChatterji2013}. 

Three further constraints can be placed on the spin and lattice modulations; detailed derivations are given in the Supplemental Material~\cite{SM}.

First, the $\kTh$ structural modulation is an internal lattice distortion with a longitudinal $c$-axis displacement component. In $R\bar{3}m$, nuclear $(0,\,0,\,L)$ reflections occur only for $L=3n$, so the satellites at $\pm1.35=0\pm q$ and $\pm1.65=\pm(3-q)$ sample the same modulation, with $q=1.35$~\rlu, in different Brillouin zones. Satellites are clearly resolved around $\bm G=(0,\,0,\,3)$, whereas no corresponding excess with statistical significance is detected around $\bm G=(0,\,0,\,0)$. Because the sum of the two polarization channels cancels the interference term, a constrained analysis of the polarization-summed integrated intensity, $I\,(L)\equiv I^{+}(L)+I^{-}(L)$, gives $I\,(1.65)/I\,(1.35)>18$, with a conservative lower bound of $13$ after inflating the fitting uncertainties~\cite{SM}. The measured asymmetry further requires the nuclear contribution to account for at least $21\%$ of the total satellite intensity at $L=1.65$. Thus, even if the upper limit at $L=1.35$ were entirely nuclear, the nuclear-intensity contrast would exceed $3.9$ for the nominal uncertainties and $2.5$ under the conservative error inflation. For a rigid displacement of the crystallographic basis, $F_N(\bm G+\bm q)\propto iQ\,u_cF_{\mathrm{nuc}}(\bm G)$. The refined parent structure gives $F_{\mathrm{nuc}}(0)=793.8$~fm and $|F_{\mathrm{nuc}}(003)|\simeq66$~fm~\cite{Zhai2017}. Including the $Q^2$ prefactor, this model would make the satellite around $\bm G=(0,\,0,\,0)$ approximately 100 times stronger than that around $\bm G=(0,\,0,\,3)$, opposite to the observed nuclear contrast. Even under the conservative error inflation, the predicted and observed intensity ratios differ by more than two orders of magnitude. The $\kTh$ response therefore cannot be described as a rigid translation of the complete crystallographic basis and must involve relative displacements within it. Since the scattering vector is parallel to $\bm c^{*}$, the nuclear satellite establishes a longitudinal $c$-axis displacement component.

Second, the interfering magnetic component is field-parallel. Neutrons measure only $\Mperp=\bm M-(\bm M\!\cdot\!\Qhat)\Qhat$. Along the $(0,\,0,\,L)$ rod, $\Qhat\parallel\bm c$, so the $c$-axis moment is projected out. Of the two remaining in-plane components, only that parallel to $\bm H\parallel\Phat$ contributes to $\Phat\!\cdot\!\mathrm{Re}(N^{*}\Mperp)$. The interference therefore isolates the field-parallel magnetic modulation along $[100]$.

Third, the nuclear and magnetic structure factors are phase-coherent. Once the conjugate-satellite comparison identifies the asymmetry with nuclear-magnetic interference, Eq.~\eqref{eq:asymmetry} gives $|r|\leq|\cos\varphi|$, where $\varphi\equiv\arg(N^{*}M_{\parallel H})$. A profile-likelihood analysis of the data in Fig.~\ref{fig:2}(c), with a common $\cos\varphi$ across temperature and independently profiled nuclear and magnetic amplitudes, gives $|\varphi|\lesssim$35$^{\circ}$. Thus, the relative nuclear-magnetic structure-factor phase remains well-defined over the measured temperature range~\cite{SM}. 

We now discuss whether the lattice modulation is magnetically induced or pre-exists based on three observations. (i)~Fitting $I^{\pm}=|N|^{2}+|M|^{2}\pm2|N||M|\cos\varphi$ to   the $7$~T temperature series, with $M(T)$ free at each temperature and $\cos\varphi$ common, a nuclear amplitude proportional to the magnetic one fits poorly ($\chi^{2}=45$ for $21$ degrees of freedom), a temperature-independent one fits better ($\chi^{2}=30$), and a power law $N\propto M^{p}$ is best at $p=0.2(1)$ ($\chi^{2}=25$ for $20$)~\cite{SM}. The data thus mildly prefer a nuclear amplitude that falls much more slowly with temperature than the magnetic   one, the nuclear fraction rising from $0.2$ at $15$~K to about $0.7$ at $275$~K, although the preference does not survive the twofold error inflation adopted   elsewhere. (ii)~In zero field the polarization-summed intensity at $L=1.65$ is consistent with zero, with a $95\%$ upper limit of $0.3$ times its $7$~T   value~\cite{SM}. A pre-existing structural modulation would retain its nuclear fraction $f_N$ of that intensity at zero field; with $0.21\leq f_N\leq0.79$   admissible, this alternative is excluded for $f_N\gtrsim0.3$ but not for the smallest nuclear fractions. (iii)~The $\kTh$ linewidth narrows from $0.22(2)$ to   $0.176(5)$~\rlu\ between $2$-$3$ and $4$-$5$~T, a $2\sigma$ effect~\cite{SM}, as expected for a correlation length that grows with field but not for a   static intergrowth period. Together these favour a field-induced nuclear modulation that is more robust against temperature than the accompanying magnetic   amplitude, consistent with a distortion stabilized by the saturated uniform moment $M_{0}$, which is nearly temperature independent below $275$~K, rather than   by the modulated moment alone; a pre-existing structural modulation locked to the field-polarized spins is disfavoured but not excluded.

Taken together, these constraints establish phase-coherent, field-parallel magnetic and internal longitudinal $c$-axis lattice modulations at the same incommensurate wavevector. Because the coherent $\kTh$ state is stabilized by an in-plane magnetic field, the relevant parent symmetry is the field-reduced subgroup of $R\bar{3}m1'$. For $\bm H\parallel[100]$, the field-reduced point group is $2/m$, and for $\bm q\parallel\bm c$ the corresponding little co-group is $m$. The longitudinal displacement $u_c(\bm q)$ and the observed field-parallel magnetic component $M_a(\bm q)$ transform according to the same spatial representation, $\Lambda_1$ \cite{SM}. They differ under time reversal, however: $u_c$ is time-even, whereas $M_a$ is time-odd. Their lowest-order coupling therefore requires the uniform field-induced moment $M_0$ and has the form
\begin{equation}
\mathcal{F} = i\lambda M_{0} \left( u_{c}^{*}M_{a}-u_{c}M_{a}^{*}
\right).
\label{eq:flock}
\end{equation}
This invariant vanishes when $M_0=0$ and provides a symmetry-allowed field-enabled magneto-elastic channel at the fundamental wave vector. It is consistent with a coarse-grained exchange-striction mechanism containing an $M_0M_{\bm q}$ cross term, although the present data do not determine the microscopic coupling.

Our data do not determine whether the structural modulation is inversion-breaking or carries a nonzero mode effective charge: neutron scattering weights the displacements by coherent scattering lengths, whereas the polarization weights them by Born effective charges, and centrosymmetric nonpolar modes are compatible with the observed structure factors~\cite{SM}. Establishing a polarization density wave, which the field-reduced symmetry permits, requires the displacement eigenvector and its mode effective charge.

In summary, we provide evidence for a distinct high-field magnetoelastic state in \BSMFO, characterized by phase-coherent magnetic and internal longitudinal $c$-axis lattice modulations at the same fundamental incommensurate wavevector. The even polarization asymmetry of the two conjugate satellites identifies nuclear--magnetic interference, while the zone dependence excludes a rigid translation of the crystallographic basis and establishes an internal mode involving relative atomic displacements. These measurements directly establish coherence between the nuclear and field-parallel magnetic structure factors above $2$~T and at $5$~K. The similarly large asymmetry observed up to $275$~K is consistent with robust interference up to room temperature. The results reveal fundamental-wavevector magnetoelastic order in a regime where the conventional single-$k$, zero-uniform-moment selection rule would place the leading structural response at $2k$. More generally, the even-odd parity comparison at conjugate satellites provides a phase-sensitive test of the nuclear-magnetic origin of polarization asymmetries in incommensurate magnets.

\begin{acknowledgments}
O.S. acknowledges the financial support from the Swedish Foundation for Strategic Research (SSF) through SwedNess (grant no. GS1715-0008). A.S. acknowledges funding from the Tillväxtverket European Regional Development Fund project SREss3 (2020-2023). E.B. and L.S. acknowledge support from the Swedish Research Council (Vetenskapsrådet) under Project Nos. 2018-04704 and 2021-06157. The 17 T magnet used on ZOOM was funded by the UK Engineering and Physical Sciences Research Council grant number EP/J016977/1 \cite{Holmes2012}. The authors gratefully acknowledge the Institut Laue-Langevin and the ISIS Neutron and Muon Source for the allocated beamtime (RB2210358). 

\end{acknowledgments}


\bibliography{refs}

\end{document}